# Observation of Anomalous Phonons in E-Type Orthorhombic $RE$MnO$_3$

P. Gao[1,a], H.Y. Chen[1], T. A. Tyson[1,2,a], Z. X. Liu[3], J. M. Bai[4], L. P. Wang[5], Y. J. Choi[2] and S.-W. Cheong[2]

1 Department of Physics, New Jersey Institute of Technology, Newark, NJ 07102
2 Rutgers Center for Emergent Materials and Department of Physics and Astronomy, Rutgers University, Piscataway, NJ 08854
3 Geophysical Laboratory, Carnegie Institution of Washington DC 20015
4 Oak Ridge National Laboratory Oak Ridge, TN 37831
5 Mineral Physics Institute, Stony Brook University, Stony Brook, NY 11794

We observe the appearance of a phonon near the lock-in temperature in orthorhombic $RE$MnO$_3$ ($RE$: Lu and Ho) and anomalous phonon hardening in orthorhombic LuMnO$_3$. The anomalous phonon occurs at the onset of spontaneous polarization. No such changes were found in incommensurate orthorhombic DyMnO$_3$. These observations directly reveal different electric polarization mechanisms in the E-type and IC-type $RE$MnO$_3$.

[a]Authors to whom correspondence should be addresses:

pg35@njit.edu and tyson@adm.njit.edu

*RE*MnO$_3$ (*RE*: Rear-earth) manganites are found to exhibit simultaneous ferroelectric and magnetic properties (multiferroic) and they have potential in spintronic applications [1]. Two types of structures are naturally formed for *RE*MnO$_3$: orthorhombic (*RE*: La-Dy) and hexagonal (*RE*: Sc, Y, Ho-Lu) structures [2] depending on the ionic radius of the rear-earth elements. By using high temperature and high pressure synthesis [3], film deposition [4] and chemical solution [5] methods, the hexagonal structure of the small ionic radius systems can be stabilized in a metastable orthorhombic (O) (P*bnm* space group) phase.

Metastable O-*RE*MnO$_3$ exhibits E-type antiferromagnetic (AFM) order in contrast with the A-type AFM (La to Sm) and incommensurate (IC) cycloidal spin structures (Dy and Tb [6, 7, 8]). For all O-*RE*MnO$_3$, recent measurements of polarization yield values: YMnO$_3$ (250 µC/m$^2$ at 4.5 K, 0 T), HoMnO$_3$ (80 µC/m$^2$ at 4.5 K, 0 T) [9], DyMnO$_3$ [10] (~ 2000 µC/m$^2$ at 10K) and TbMnO$_3$ [11] (800 µC/m$^2$ at 10K). These modest values should be compared with the value of 7.5x10$^5$ µC/m$^2$ for PbTiO$_3$, whose ferroelectricity is related to off center displacements of Ti and Pb by ~0.2 Å [12]. In the O-*RE*MnO$_3$ systems, it is predicted to be mainly due to changes in the electron charge distributions. The E-type O-*RE*MnO$_3$ have been predicted to exhibit mainly electric polarization [13,14] and the large spontaneous polarization (P$_{HoMnO3}$ = 6 x 10$^4$ µC/ m$^2$) is along the *a*-axis (P*bnm*) in a combined model (atomic and electronic degrees of freedom) [15,16,17]. However, they are difficult to prepare as single crystals. Probing the properties by multiple experimental methods will provide a clearer picture of the nature of the electric polarization by specifying its true origin (electronic and/or atomic) and possibly provide paths for its enhancement and its coupling with magnetic fields.

Here, we apply infrared (IR) absorption spectroscopy to study the phonon properties of O-$RE$MnO$_3$ ($RE$: Lu, Ho and Dy) in a large temperature range, which covers the AFM Néel temperature ($T_N$) and the lock-in temperature ($T_L$, the magnetic propagation vector remains locked at a constant value below this temperature [11]). Polycrystalline O-LuMnO$_3$ and O-HoMnO$_3$ were synthesized at 6 GPa and 1300 °C in a 2000-ton split-sphere multi-anvil apparatus (USSA-2000) at Stony Brook University. Duration at peak pressure and temperature was 1 hour. X-ray powder diffraction shows no residual hexagonal phase. The temperature dependence of magnetization (Fig. 1), where $T_N$ = 42 K is evidenced, was measured in a SQUID magnetometer (Quantum Design MPMS) upon warming in H = 2 T after cooling in zero magnetic field. The temperature dependent electric polarization (Fig. 1) was obtained by the integration of pyroelectric current on a poled ceramic, measured in a Quantum Design PPMS using an electrometer with the temperature variation of 4 K/min. The polarization starts to appear below $T_L$ = 36 K and its magnitude reaches ~13 µC/m$^2$ at 2 K. O-DyMnO$_3$ was synthesized by the solid-state reaction method ($T_N$ ~ 39 K and $T_L$ ~19 K [10]). O-$RE$MnO$_3$ powders were mixed with CsI by weight ratios of 1: 50 and 1: 200 for IR measurements of weak and strong absorption features, respectively. The pellet thickness was 255 µm and the sample was placed on a cryostat cold finger. The lowest temperature reached was 13 K. IR absorption spectra were collected at National Synchrotron Light Source (NSLS) beam line U2A, with a resolution of 4 cm$^{-1}$, in the range of 100 – 650 cm$^{-1}$.

O-LuMnO$_3$ has several interesting properties to compare with the other two samples: the small radius generating a very strong distortion in the orthorhombic crystal structure [18], short bond lengths, close $T_N$ (~ 41 K) and $T_L$ (~ 36 K) [19], and filled 4$f$ orbits on $Lu^{3+}$. In our measurements shown below, we observed the appearance of a phonon near $T_L$ in O-LuMnO$_3$ and O-HoMnO$_3$, but not in O-DyMnO$_3$. Fig. 2 (a), (b) and (c) are the high weight ratio (1:50) IR absorption spectra for O-LuMnO$_3$, O-HoMnO$_3$ and O-DyMnO$_3$. As expected, phonons harden with the temperature decrease. The high weight ratio spectra of REMnO$_3$ display a smooth background and clear absorption peaks below 400 cm$^{-1}$. In the region between 400 cm$^{-1}$ and 450 cm$^{-1}$, the strong absorption peaks are saturated. Phonons in the saturation band were checked in the 1:200 weight ratio spectra. The O-LuMnO$_3$ sample was measured in a cooling and warming process in the temperature range 300 K to 13 K. During the cooling process, a phonon (~ 480 cm$^{-1}$, small peak) appears in the spectrum measured at 33 K (near $T_L$= 36 K) and persists at low temperatures (see Fig. 2 (a)). This phonon disappears when the temperature is raised above $T_L$ again (see the spectrum measured at 39K). In O-HoMnO$_3$ also, a peak (~ 470.5 cm$^{-1}$, but weaker) is observed to start at ~25 K ($T_L$ ~26 K and $T_N$ ~42 K [9]) and disappears above $T_L$ in the IR spectra Fig. 2(b). The result of O-HoMnO$_3$ again confirms that the appearance of a phonon is at the lock-in transition temperature. In contrast, in Fig. 2(c), no peak appears in the absorption spectra of IC O-DyMnO$_3$.

From symmetry analysis, 25 IR sensitive and 24 Raman active modes are predicted (four formula units per unit cell). In the energy range 450 ~ 650 cm$^{-1}$, Raman phonons arise from O2 stretching motions and the bending motion of the MnO$_6$ polyhedra. The phonon (~ 480 cm$^{-1}$) is close to the $B_{2g}(3)$ Raman phonon of HoMnO$_3$ and YMnO$_3$ (481cm$^{-1}$) [20]. This Raman phonon is related to out-of-plane MnO$_6$ bending and the IR phonon would have a similar behavior even though the phonons are not in a pure bending or stretching mode. In the strongly distorted P*bnm* cell, the displacement of O is much larger than those of *R* and Mn. The high frequencies (~ 600 cm$^{-1}$) correspond to the oxygen dominated modes. Those strong O2 displacements favor the ferroelectricity prediction of E-phase orthorhombic perovskite [15]. Comparing the structures, the lattice of DyMnO$_3$ is more ordered and compact than the lattice of LuMnO$_3$, hence at low temperature, the effect of tilting of the MnO$_6$ in O-LuMnO$_3$ may lead to a lower symmetry which makes the IR phonon (~480cm$^{-1}$) visible. With the modified lower symmetry structure, freedom is added and the Mn atom can lose its inversion center. This lowering of the symmetry leading to minor displacements, enables one to understand the anomalous low temperature enhancement of the atomic displacement parameters in neutron diffraction data of O-LuMnO$_3$, whose *B* values at 300K (8K) are 0.34 (0.44) Å$^2$ for Mn, 0.17 (0.32 Å$^2$) for O1 and 0.17 (0.32) Å$^2$ for O2 [21].

At 300 K, 14 phonons are found for O-LuMnO$_3$ below 600 cm$^{-1}$ by a Lorentz model least square fit. For O-LuMnO$_3$, the phonon at 335.7 cm$^{-1}$ anomalous behavior is shown in Fig. 3. Below ~ 38 K (close to T$_L$), the hardening rate of this phonon has a sudden change. This phonon (335.7 cm$^{-1}$) is one characteristic frequency of the MnO$_6$ polyhedra, which involves complex motions of the Mn atoms and the O atoms [22, 23] along *c*-axis (B$_{1u}$ mode). This frequency doesn't depend on the *RE* ion size and it can be found in TbMnO$_3$ (335.4 cm$^{-1}$) [24, 25] and LaMnO$_3$ (335 cm$^{-1}$) [23]. When the temperature drops, this phonon hardens and the atomic motion is restricted along the *c*-axis. In this mode, the possible motions of Mn and O are confined to the *ab* plane. This constraint on atomic displacements to the *ab* plane is consistent with the theoretically predicted [15] spontaneous polarization along *a*-axis.  In a contrast, the $B_{1u}$ phonon (335cm$^{-1}$ here) in IC DyMnO$_3$ along *c*-axis (335.4 cm$^{-1}$ in TbMnO$_3$ [24, 25]) does not exhibit anomalous hardening.  In O-HoMnO$_3$ absorption spectrum, the phonon 335cm$^{-1}$ can be observed, but doesn't show the anomalous hardening like in O-LuMnO$_3$.  We think that the onset of E-type magnetic ordering results in a lowering of symmetry which is stronger in the Lu system (with smaller radii) than in the Ho system.  Hence the onset of the distortion is manifested by a change in the phonon frequency of the 335 cm$^{-1}$ mode in O-LuMnO$_3$.

Over all, the appearance of phonons (*RE* = Ho at ~ 470.5cm$^{-1}$ and Lu at ~ 480 cm$^{-1}$) and anomalous phonon (~ 335.7 cm$^{-1}$) hardening (in *RE* = Lu) were observed in E-type O-*RE*MnO$_3$ near $T_L$.  Anomalous phonon behaviors imply a different spontaneous polarization mechanism for E-type O-*RE*MnO$_3$ in contrast with the IC O-*RE*MnO$_3$, where the complex spiral spin order leads to broken spatial inversion symmetry.  In E-type systems, the onset of magnetic ordering leads to weak structural distortions [21] producing  a splitting of phonon modes and phonon hardening.  In these orthorhombic materials (IC and E type), magnetic ordering initiates a charge rearrangement and weak atomic displacements yielding a finite electric polarization. The enhancement of the magnetic coupling to the lattice may yield systems with larger electrical polarization.

This research was funded by DOE Grants DE-FG02-07ER46402 (NJIT) and DE-FG02-07ER46382 (Rutgers).  The U2A beam line at the National Synchrotron Light Source is supported by COMPRES, the Consortium for Materials Properties Research in Earth Sciences under NSF Cooperative Agreement EAR01-35554, U.S. Department of Energy (DOE-BES and NNSA/CDAC).  Use of NSLS at Brookhaven National

Laboratory, was supported by the U.S. Department of Energy, Office of Science, Office of Basic Energy Sciences, under Contract No. DE-AC02-98CH10886.

# Figure Captions

**Figure 1.** (Color online) Electric polarization (H = 0 T) and magnetization (H = 2 T) of O-LuMnO$_3$

**Figure 2.** (Color online) (a) Temperature dependent infrared absorption spectra of O-LuMnO$_3$. Thin lines are for the cooling process and thick lines are for the warming process. Blue and red arrows indicate the anomalous phonon and the appearance of a phonon. (b) Infrared absorption spectra of O-HoMnO$_3$. (c) Infrared absorption spectra of O-DyMnO$_3$. The break is from 410 cm$^{-1}$ - 440 cm$^{-1}$, where the phonon band absorption is strong (also a region of strong absorption is excluded for some spectra in O-HoMnO$_3$ near 550 cm$^{-1}$). The inset is the magnification of the phonon for the range 440 cm$^{-1}$- 500 cm$^{-1}$.

**Figure 3.** The anomalous phonon hardening of O-LuMnO3 (335.7 cm$^{-1}$). The vertical line indicates the onset temperature near T$_L$.

**Fig. 1. P. Gao *et al.***

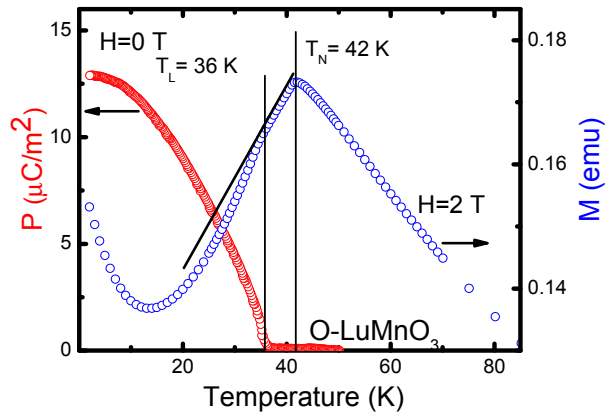

**Fig.2. P. Gao *et al.***

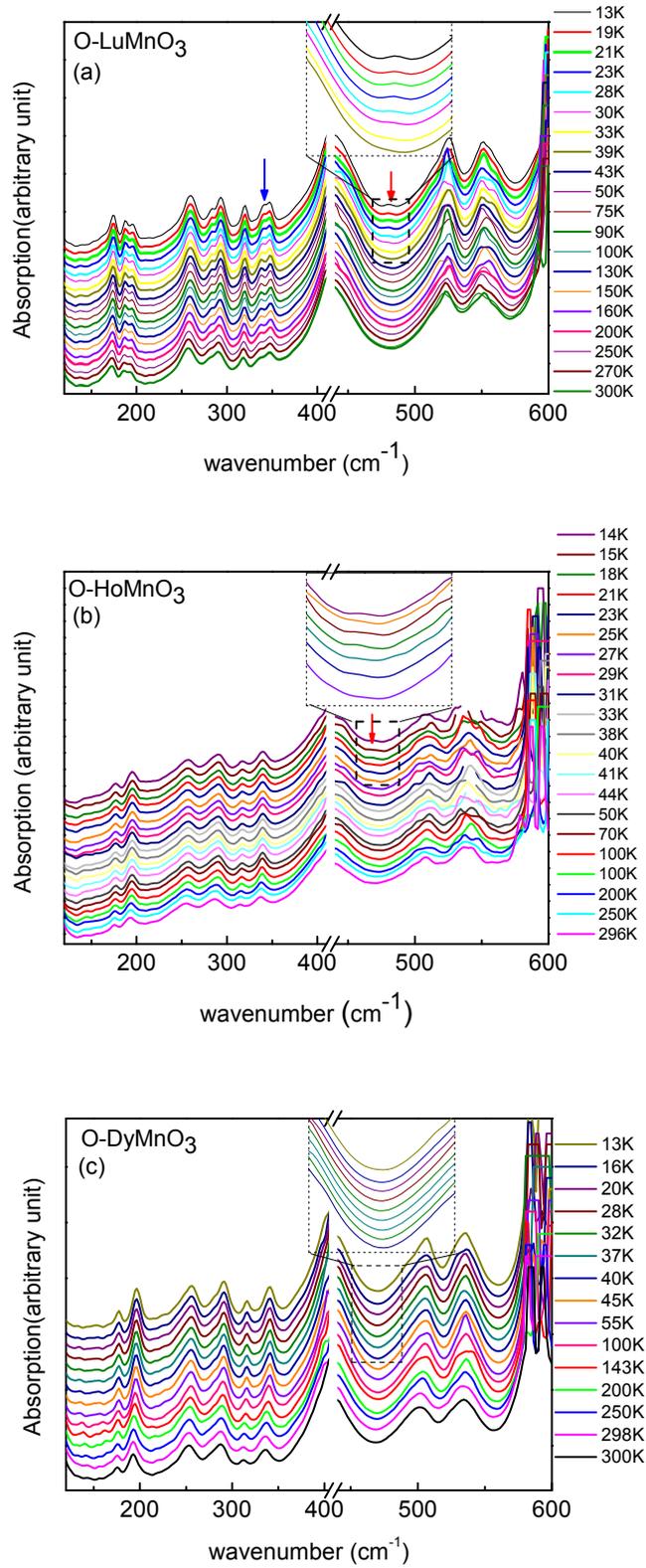

**Fig. 3. P. Gao *et al.***

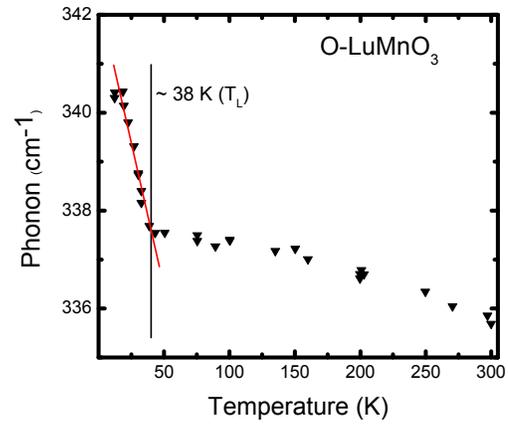